# Anomalous Hall Conductivity of a Non-Collinear Magnetic Antiperovskite

Gautam Gurung, Ding-Fu Shao, Tula R. Paudel, and Evgeny Y. Tsymbal

*Department of Physics and Astronomy & Nebraska Center for Materials and Nanoscience,*
*University of Nebraska, Lincoln, Nebraska 68588-0299, USA*

The anomalous Hall effect (AHE) is a well-known fundamental property of ferromagnetic metals, commonly associated with the presence of a net magnetization. Recently, an AHE has been discovered in non-collinear antiferromagnetic (AFM) metals. Driven by non-vanishing Berry curvature of AFM materials with certain magnetic space group symmetry, anomalous Hall conductivity (AHC) is very sensitive to the specific type of magnetic ordering. Here, we investigate the appearance of AHC in antiperovskite GaNMn$_3$ as a representative of broader materials family ANMn$_3$ (A is a main group element), where different types of non-collinear magnetic ordering can emerge. Using symmetry analyses and first-principles density-functional theory calculations, we show that with almost identical band structure, the nearly degenerate non-collinear AFM $\Gamma_{5g}$ and $\Gamma_{4g}$ phases of GaNMn$_3$ have zero and finite AHC, respectively. In a non-collinear ferrimagnetic M-1 phase, GaNMn$_3$ exhibits a large AHC due to the presence of a sizable net magnetic moment. In the non-collinear antiperovskite magnets, transitions between different magnetic phases, exhibiting different AHC states, can be produced by doping, strain, or spin transfer torque, which makes these materials promising for novel spintronic applications.

## I. Introduction

It is known that the anomalous Hall effect (AHE) emerges in metals with broken time-reversal symmetry (TRS) and strong spin-orbit coupling (SOC) [1]. Usually, the AHE is found in ferromagnetic (FM) metals, where a transverse voltage generated by a longitudinal charge current is sensitive to the net magnetization. The intrinsic AHE is driven by a fictitious magnetic field in the momentum space associated with the Berry curvature, a quantity inherent in the electronic band structure [2]. With the magnitude and direction determined by the magnetization and SOC, this fictitious magnetic field controls the charge current in a similar way as a real magnetic field in the ordinary Hall effect. The AHE vanishes in conventional collinear antiferromagnetic (AFM) metals due to the anomalous Hall conductivities being opposite in sign and hence cancelling each other for the two ferromagnetic sublattices with opposite magnetization. In other words, the existence of symmetry combining time reversal and lattice translation prohibits the AHE. This observation suggested that the presence of a non-vanishing net magnetic moment is the necessary condition to break the related symmetry and produce the AHE [3].

It appeared, however, that the AHE can be observed in certain types of non-collinear antiferromagnets, such as Mn$_3$X alloys (X = Ga, Ge, Ir, etc.) [4–8]. In these metals, the Mn moments are arranged in a Kagome-type lattice within the (111) plane. The magnetic space group symmetry operations in these compounds cannot eliminate the total Berry curvature, leading to a non-vanishing AHE [4]. The presence of a sizable AHE in non-collinear AFM metals is interesting for AFM spintronics, where an AFM order parameter, as a state variable, can be controlled on a much shorter time scale compared to magnetization in ferromagnets [9–11].

Importantly, specifics of magnetic ordering in non-collinear AFM materials associated with different magnetic space group symmetries have a strong impact on the AHE [12, 13]. For example, it was found that the AHC tensors have a different form in Mn$_3$X (X = Ga, Ge, and Sn) and Mn$_3$Y (Y = Rh, Ir, and Pt) compounds, due to different magnetic moment configurations. One can expect therefore that a significant change in the anomalous Hall conductivity (AHC) can emerge at the magnetic phase transition associated with switching between different non-collinear magnetic orderings. Realizing such an effect in practice would be interesting for potential spintronic applications, and therefore exploring the AHE in possible material systems with competing and tunable non-collinear magnetic phases is valuable.

Antiperovskite materials are potential candidates for the control of the AHE by tunable non-collinear magnetism. Antiperovskites have a perovskite structure, where cation and anion positions are interchanged (Fig. 1(a)). Abundant functional properties have been discovered in these materials, such as superconductivity [14], magnetoresistance [15], and magnetovolume [16–18], magnetocaloric [19, 20], and barocaloric [21] effects. Manganese nitride antiperovskites ANMn$_3$ (A = Ga, Cu, Ni, etc.) are typically metallic and often reveal complex magnetic orderings [16, 22, 23]. Various magnetic phases, such as non-collinear AFM $\Gamma_{5g}$ and $\Gamma_{4g}$ phases and a non-collinear ferrimagnetic M-1 phase have been found in these compounds (Fig. 1). Transformations between these magnetic phases can be induced by perturbations, such as doping, pressure, and temperature [23–25]. It has also been predicted that the transition between the $\Gamma_{5g}$ and $\Gamma_{4g}$ phases can be achieved using a spin transfer torque [26]. These properties make ANMn$_3$ compounds promising for a functional control of the non-collinear magnetism and thus interesting for exploring the AHE in different magnetic phases.

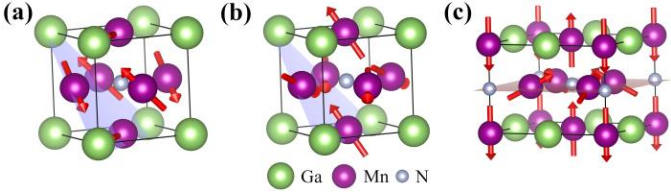

**Figure 1:** Different non-collinear magnetic phases in AFM antiperovskite GaNMn$_3$: (a) $\Gamma_{5g}$, (b) $\Gamma_{4g}$, and (c) M-1. Red arrows denote magnetic moments.

In this paper, we consider gallium manganese nitride GaNMn$_3$ as a representative antiperovskite material to investigate the magnetic phase dependent AHC of the whole ANMn$_3$ family. The high temperature paramagnetic phase of GaNMn$_3$ has a cubic crystal structure with the space group $Pm\bar{3}m$. The $\Gamma_{5g}$ phase emerges below room temperature (Fig. 1(a)) and represents the most common non-collinear AFM phase of the ANMn$_3$ compounds. In this phase, to avoid the frustration from the triangular geometry of the Ga-Mn Kagome-type lattice in the (111) plane, the magnetic moments of the three Mn atoms form a chiral configuration with the 120° angle between each other. The $\Gamma_{4g}$ magnetic structure is another common non-collinear AFM phase in the ANMn$_3$ family, which can be obtained from the $\Gamma_{5g}$ phase by rotating all magnetic moments around the [111] axis by 90° (Fig. 1(b)). Both the $\Gamma_{5g}$ and $\Gamma_{4g}$ phases have zero net magnetization. GaNMn$_3$ also exhibits a non-collinear ferrimagnetic M-1 phase (Fig. 1(c)), which can be stabilized by stoichiometric deficiency or high pressure [23]. In this phase, the Mn magnetic moments are antiferromagnetically (ferromagnetically) coupled in (between) the Ga-Mn (001) planes, resulting in collinear AFM sublattices within these planes. On the other hand, the magnetic moments in the Mn-N (002) planes are arranged non-collinearly (Fig. 1(c)), leading to the net magnetic moment along the [001] direction.

Using symmetry analyses and first-principles density-functional theory (DFT) calculations, we explore the AHE of the three non-collinear magnetic phases of GaNMn$_3$. We show that with nearly identical band structure, the nearly degenerate AFM $\Gamma_{5g}$ and $\Gamma_{4g}$ phases have zero and finite AHC, respectively. In a non-collinear ferrimagnetic M-1 phase, GaNMn$_3$ exhibits a large AHC due to the presence of a sizeable net magnetization. With a possibility to control the appearance of these magnetic phases by external stimulus, the predicted variation of the AHC between different magnetic phases in the same material point to a new approach of designing the AHE-based functional devices for spintronic applications.

## II. Symmetry analysis

Within the linear response theory, the intrinsic AHC is expressed as the integral of the total Berry curvature ($\Omega_{\alpha\beta}$) over the Brillouin zone of the crystal [1, 27]

$$\sigma_{\alpha\beta} = -\frac{e^2}{\hbar}\oint_{BZ}\frac{d^3\vec{k}}{(2\pi)^3}\Omega_{\alpha\beta}(\vec{k}), \quad (1)$$

where the total Berry curvature $\Omega_{\alpha\beta} \equiv \Omega^\gamma = \sum_n f_n(\vec{k})\Omega_n^\gamma(\vec{k})$ is the sum of the Berry curvatures $\Omega_n^\gamma(\vec{k})$ corresponding the individual bands $n$, $f_n(\vec{k})$ is the Fermi distribution function, and indices ($\alpha$, $\beta$, $\gamma$) denote Cartesian co-ordinates. The expression for the Berry curvature $\Omega_n^\gamma(\vec{k})$ is given by [1, 27]

$$\Omega_n^\gamma(\vec{k}) = -2i\hbar^2 \sum_{m\neq n} \frac{\langle\psi_{n,\vec{k}}|v_\alpha|\psi_{m,\vec{k}}\rangle\langle\psi_{m,\vec{k}}|v_\beta|\psi_{n,\vec{k}}\rangle}{(E_m(\vec{k})-E_n(\vec{k}))^2}, \quad (2)$$

where $\psi_{n,\vec{k}}$ is the Bloch function and $\vec{v}$ is the velocity operator. Space group symmetry of a material determines the presence or absence of a finite AHC. For example, since $\Omega_n^\gamma(\vec{k})$ is odd with respect to time reversal symmetry, i.e. $\Omega_n^\gamma(-\vec{k}) = -\Omega_n^\gamma(\vec{k})$, the total Berry curvature $\Omega_{\alpha\beta}$ and hence the AHC are zero for non-magnetic materials. Similarly, if there is symmetry operation $\hat{O}$ transforming $\vec{k}$ to $\vec{k}'$ (i.e., $\vec{k}' = \hat{O}\vec{k}$), such as two-fold rotation or mirror reflection, for which $\hat{O}\Omega_n(\vec{k}') = -\Omega_n(\vec{k})$, the AHC vanishes [12, 13]. In non-collinear AFM materials, such as GaNMn$_3$, various magnetic phases are associated with different magnetic space group symmetries, resulting in different AHC.

The $\Gamma_{5g}$ phase of GaNMn$_3$ is characterized by a lattice of magnetic "whirls" composed of non-collinear Mn magnetic moments in the (111) plane (Fig. 2(a)). This arrangement forms the magnetic space group $R\bar{3}m$, which has three mirrors planes perpendicular to the (111) plane. Application of the symmetry transformations $M = M_{0\bar{1}1}$, $M_{10\bar{1}}$, or $M_{\bar{1}10}$ preserves the original configuration of magnetic moments. The invariance under the mirror symmetry transformations causes the AHE in the $\Gamma_{5g}$ phase to vanish. For example, under the $M_{\bar{1}10}$ symmetry operation, the Berry curvature is transformed as $\Omega_{xy}(k_x, k_y, k_z) = -\Omega_{xy}(k_y, k_x, k_z)$, which implies that the integral over the whole Brillouin zone in Eq. (1) is zero.

This odd property of the Berry curvature in GaNMn$_3$ under the mirror symmetry transformations is broken in the $\Gamma_{4g}$ phase. In this phase, the Mn magnetic moments form a lattice of "vertices" in the (111) plane (Fig. 2(b)). This configuration corresponds to the magnetic space group $R\bar{3}m'$, in which the mirror symmetries are broken. As seen from Fig. 2(b), mirror symmetry transformation $M$ reverses all the magnetic moments.

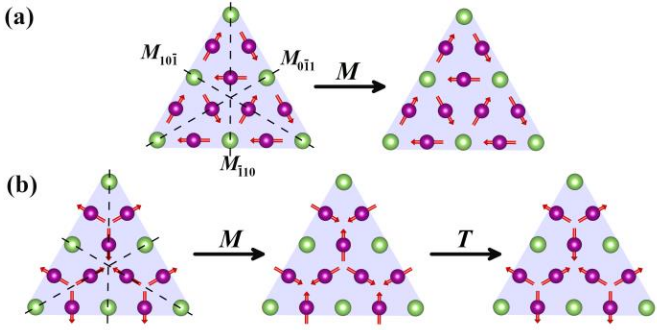

**Figure 2:** Symmetry operations for non-collinear AFM phases $\Gamma_{5g}$ (a) and $\Gamma_{4g}$ (b) in the (111) Ga-Mn plane of GaNMn$_3$. (a) The $\Gamma_{5g}$ phase preserves mirror planes $(\bar{1}10)$, $(10\bar{1})$ and $(0\bar{1}1)$ (denoted by dashed lines) and is invariant under symmetry transformations $M = M_{0\bar{1}1}, M_{10\bar{1}}$ or $M_{\bar{1}10}$. (b) The $\Gamma_{4g}$ phase does not preserve the mirror planes, but is invariant under the product of mirror symmetry M and time reversal symmetry T. Red arrows denote the magnetic moments. Dotted lines denote the mirror planes.

**Table 1:** Matrix elements of the AHC tensor for different magnetic phases in GaNMn$_3$. Here, the ordinary Cartesian coordinates are used, i.e. $\hat{x}||[100]$, $\hat{y}||[010]$, and $\hat{z}||[001]$.

| Magnetic Phase | $\Gamma_{5g}$ | $\Gamma_{4g}$ | M-1 |
|---|---|---|---|
| Magnetic Space Group | $R\bar{3}m$ | $R\bar{3}m'$ | P4 |
| AHC tensor | $\begin{bmatrix} 0 & 0 & 0 \\ 0 & 0 & 0 \\ 0 & 0 & 0 \end{bmatrix}$ | $\begin{bmatrix} 0 & \sigma_{xy} & -\sigma_{xy} \\ -\sigma_{xy} & 0 & \sigma_{xy} \\ \sigma_{xy} & -\sigma_{xy} & 0 \end{bmatrix}$ | $\begin{bmatrix} 0 & \sigma_{xy} & 0 \\ -\sigma_{xy} & 0 & 0 \\ 0 & 0 & 0 \end{bmatrix}$ |

In contrast, the product of mirror symmetry $M$ and time reversal symmetry $T$ is preserved in the $\Gamma_{4g}$ phase. As shown in Fig. 2(b), when reversal of all moments by the mirror symmetry operation $M$ is followed by the time reversal symmetry transformation $T$ all the moments are reversed back to their initial configuration. The presence of the combined $TM$ symmetry makes the Berry curvature an even function of wave vector $\vec{k}$. For example, applying the $TM_{\bar{1}10}$ transformation we obtain $\Omega_{xy}(k_x, k_y, k_z) = \Omega_{xy}(-k_y, -k_x, -k_z)$. This even property of the Berry curvature makes the AHC non-zero in the $\Gamma_{4g}$ phase.

Magnetic space group symmetry determines the shape of the AHC tensor. While in the $\Gamma_{5g}$ phase, all the nine components of the AHC tensor are zero, in the $\Gamma_{4g}$ phase, corresponding to the magnetic space group $R\bar{3}m'$, the AHC tensor is non-zero. Table 1 shows that there are six non-vanishing matrix elements of the AHC tensor in the $\Gamma_{4g}$ phase with only one $\sigma_{xy}$ being independent.

In the non-collinear ferrimagnetic M-1 phase (Fig. 1 (c)), GaNMn$_3$ has a net magnetization along the [001] direction. Therefore, a non-zero AHC is expected in this case similar to that in ferromagnetic metals. Table 1 shows the AHC tensor for the magnetic space group symmetry P4 corresponding to the M-1 phase. Like in collinear ferromagnetic metals, the AHC tensor has two non-zero components with only one $\sigma_{xy}$ being independent.

### III. Methods

Next, we perform first-principles DFT calculations to obtain the AHC of the three non-collinear magnetic phases of GaNMn$_3$. The DFT calculations are performed using a plane-wave pseudopotential method with the fully-relativistic ultrasoft pseudopotentials [28] implemented in Quantum-ESPRESSO [29]. The exchange and correlation effects are treated within the generalized gradient approximation (GGA) [30]. We use the plane-wave cut-off energy of 52 Ry and the $k$-point mesh of 16 × 16 × 16 for the cubic $\Gamma_{5g}$ and $\Gamma_{4g}$ phases and 12 × 12 × 16 for the tetragonal M-1 phase in GaNMn$_3$. Spin-orbit coupling is included in all the calculations.

The AHC is calculated using the PAOFLOW code [31] based on pseudo-atomic orbitals (PAO) [32,33]. Tight-binding Hamiltonians are constructed from the non-self-consistent DFT calculations with a 16 × 16 × 16 $k$-point mesh for the $\Gamma_{5g}$ and $\Gamma_{4g}$ phases and a 12 × 12 × 16 $k$-point mesh for the M-1 phase. Then, the AHC are calculated with a 48 × 48 × 48 $k$-point mesh for the $\Gamma_{5g}$ and $\Gamma_{4g}$ phases and a 46 × 46 × 48 $k$-point mesh for the M-1 phase using the adaptive broadening method.

The symmetry determined geometries of the AHC tensor are obtained using the FINDSYM code and the linear response symmetry code [34]. The figures are created using VESTA [35] and gnuplot [36].

### IV. Results

Our DFT calculations show that the lattice constant of GaNMn$_3$ is 3.869 Å, which is close to the experimental and previously calculated values [22, 23, 37, 38, 39]. We find that the $\Gamma_{5g}$ phase is the ground state of GaNMn$_3$, while the total energies of the $\Gamma_{4g}$ and M-1 phases are higher by 0.5 meV/f.u. and 193.5 meV/f.u., respectively. This result is consistent with the experimental observations showing the appearance of the $\Gamma_{5g}$ phase in GaNMn$_3$ at low temperature [21, 22, 24]. Recently, the M-1 phase was observed at high pressure in stoichiometrically deficient samples of GaNMn$_3$ [23].

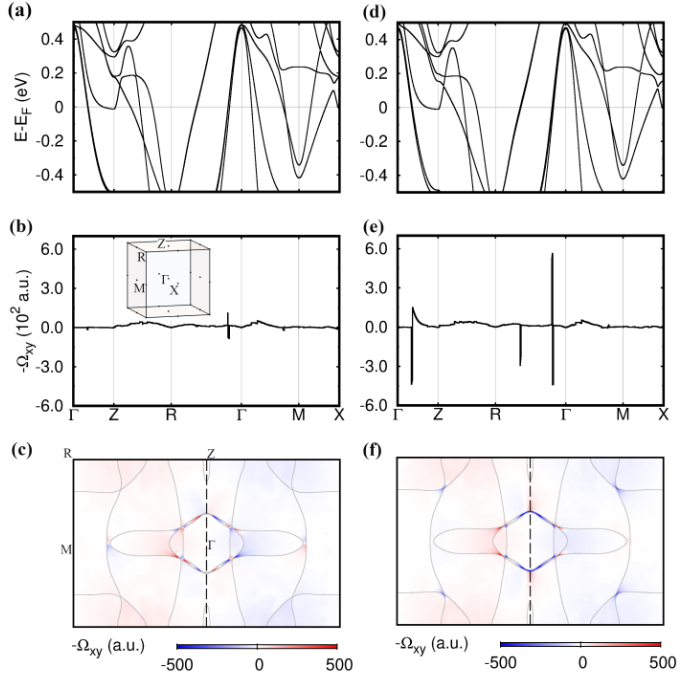

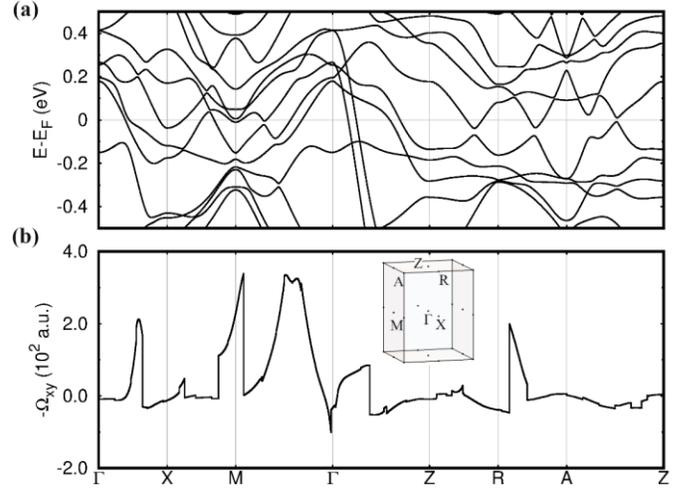

**Figure 4:** (a-c) The calculated band structure (a), Berry curvature $\Omega_{xy}$ along high symmetry path (b), and the color map of $\Omega_{xy}$ in the (110) plane (c) for the $\Gamma_{5g}$ phase of GaNMn$_3$. (d-f) The calculated band structure (d), $\Omega_{xy}$ along high symmetry path (e), and the color map of $\Omega_{xy}$ in the (110) plane (f) for the $\Gamma_{4g}$ phase of GaNMn$_3$. The inset of (b) shows the Brillouin zone. The solid lines and the dashed line in (c) and (f) denotes the Fermi surfaces and the mirror plane $M_{\bar{1}10}$.

The calculated local magnetic moment in the $\Gamma_{5g}$ and $\Gamma_{4g}$ phases is about 2.16 $\mu_B$/Mn atom, which is in a qualitative agreement with the experimental and previously calculated values [22,23,37,38]. As expected, the non-collinear AFM configuration leads to a zero net magnetic moment. For the ferrimagnetic M-1 phase, we obtain 2.10 $\mu_B$ per Mn atom in the (001) plane and 1.75 $\mu_B$ per Mn atom in the (002) plane, resulting in the net magnetic moment of 2.52 $\mu_B$/f.u pointing along the $z$ direction. These values are calculated using the lattice constant of the $\Gamma_{5g}$ phase and are slightly different from the experimental values, which is understandable due to the M-1 phase emerging only at high pressure [23].

Since $\Gamma_{5g}$ and $\Gamma_{4g}$ have similar magnetic structures, we first investigate the AHE in these two phases of GaNMn$_3$. Figure 3(a) shows the band structure of the $\Gamma_{5g}$ phase. Five bands cross the Fermi energy ($E_F$). These dispersive bands are largely composed of the Mn-3$d$ orbitals. It is seen that in some directions the bands are very close to each other. For example, along the $\Gamma$-Z and R-$\Gamma$ directions, there are nearly degenerate bands.

**Figure 3:** Calculated band structure (a) and Berry curvature $\Omega_{xy}$ (b) of GaNMn$_3$ in the M-1 phase along high symmetry paths in the Brillouin zone. The inset of (b) shows the Brillouin zone.

Figure 3(b) shows the calculated Berry curvature $\Omega_{xy}$. It is seen that there are peaks along the R-$\Gamma$ direction, which appear, according to Eq. (2), due to the small band separation between the three bands crossing $E_F$ along this direction close to the $\Gamma$ point (see Fig. 3(a)). Along the $\Gamma$-Z direction, the Berry curvature $\Omega_{xy}$ is zero within the computation accuracy. This is due to the mirror symmetry $M_{\bar{1}10}$ which holds along this high symmetry direction, resulting in $\Omega_{xy}(0,0,k_z) = -\Omega_{xy}(0,0,k_z)$, and hence $\Omega_{xy}(0,0,k_z) = 0$.

In order to demonstrate the odd nature of the Berry curvature under the mirror symmetry $M_{\bar{1}10}$, we plot in Figure 3(c) the color map of $\Omega_{xy}$ in the (110) plane, which is perpendicular to the ($\bar{1}$10) plane. It is seen that hot spots (i.e. regions where the absolute values of the Berry curvature are large) appear around the $k$-points where the Fermi surfaces of different bands (indicated by solid lines in Fig. 3(c)) cross. As is evident from Figure 3(c), $\Omega_{xy}$ changes sign with respect to the mirror symmetry transformation $M_{\bar{1}10}$ (reflection with respect to the dashed line in Fig. 3(c)). Clearly, integration of the $\Omega_{xy}$ over the whole Brillouin zone using Eq. (1) leads to zero AHC (within the computational accuracy) for the $\Gamma_{5g}$ phase. As seen from Figure 5(a), this property is independent of energy (Fermi energy).

Figure 3(d) shows the band structure of GaNMn$_3$ in the $\Gamma_{4g}$ phase. Since the $\Gamma_{4g}$ phase can be obtained from the $\Gamma_{5g}$ phase by rotation of all magnetic moments around the [111] axis by 90°, in the absence of SOC the band structure of the two phases should be identical. Thus, the subtle differences in the bands structures in Figures 3(a) and 3(d) are due to SOC. These differences are seen, particularly, along the $\Gamma$-Z and $\Gamma$-R

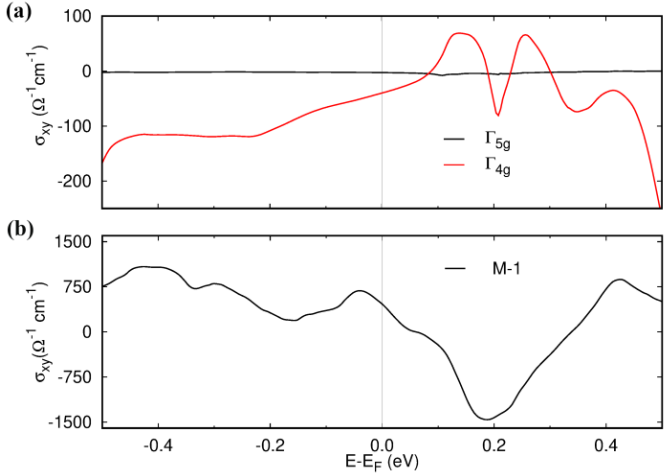

**Figure 5:** (a,b) Calculated AHC $\sigma_{xy}$ as a function of energy for the $\Gamma_{5g}$ and $\Gamma_{4g}$ (a) and M-1 (b) phases of GaNMn$_3$.

directions, where there is a slight increase in the band splitting around the Fermi energy.

Figure 3(e) shows the calculated Berry curvature of GaNMn$_3$ in the $\Gamma_{4g}$ phase and reveals pronounced peaks in $\Omega_{xy}$ along the $\Gamma$-Z and $\Gamma$-R directions. According to the $TM_{\bar{1}10}$ symmetry, $\Omega_{xy}$ is an even function of the wave vector $\vec{k}$, i.e. $\Omega_{xy}(k_x, k_y, k_z) = \Omega_{xy}(-k_y, -k_x, -k_z)$. This is reflected in the calculated color map of $\Omega_{xy}$ in the (110) plane, which is shown in Figure 3(f). It is seen that the hot spots of $\Omega_{xy}$ appear nearly at the same locations as for the $\Gamma_{5g}$ phase (Fig. 3 (c)). However, in the $\Gamma_{4g}$ phase, they are distributed symmetrically and have the same sign, proving that $\Omega_{xy}$ is an even function with respect the $TM_{\bar{1}10}$ symmetry transformation. The AHC is calculated by integration of $\Omega_{xy}$ according to Eq. (2). Figure 5(a) shows that $\sigma_{xy}$ is finite as a function of energy and at the Fermi energy $\sigma_{xy} = -40 \, \Omega^{-1} cm^{-1}$. Clearly, the difference in the AHC between the $\Gamma_{5g}$ and $\Gamma_{4g}$ phases is due to the different magnetic space group symmetry of these phases.

Figure 4(a) shows the calculated band structure of GaNMn$_3$ in the M-1 phase along high symmetry directions in the Brillouin zone. The band structure is more intricate compared to those for the $\Gamma_{5g}$ and $\Gamma_{4g}$ phases, because of a larger unit cell and more complex magnetic configuration. The presence on the net magnetic moment breaks time reversal symmetry, which makes the AHC non-zero. Figure 5(b) shows the calculated Berry curvature $\Omega_{xy}$ along the high symmetry directions. It is seen that there are a number of pronounced broad peaks which are associated with the multiple low dispersive bands around the Fermi energy which are coupled by the spin-orbit interaction. Figure 5(b) shows the calculated AHC as a function of energy in the M-1 phase. At the Fermi energy, $\sigma_{xy} = 454 \, \Omega^{-1} cm^{-1}$ which is much larger than the AHC in the $\Gamma_{4g}$ phase, due to the presence of the net magnetic moment in the M-1 phase. It is notable that $\sigma_{xy}$ can be strongly enhanced in the M-1 phase by hole doping. For example, at $E = E_F - 0.04$ eV, the calculated value of $\sigma_{xy}$ is as large as 685 $\Omega^{-1} cm^{-1}$ which is comparable to the AHC in Fe ($\sigma_{xy} \sim 700 \, \Omega^{-1} cm^{-1}$ [40]).

## V. Discussion

Our results demonstrate that in the family of antiperovskite compounds, as represented by GaNMn$_3$, the AHC is strongly dependent on the specific magnetic configuration. A significant change in the AHC can be produced by transitions between different magnetic phases. Such transitions can be driven by an external stimulus, provided that the energies of the different non-collinear magnetic phases are engineered to be nearly degenerate.

In experiment, the $\Gamma_{5g}$ phase is found in the ANMn$_3$ compounds with A = Zn, Ga, and the $\Gamma_{4g}$ phase is found for A = Ni, Ag, Sn [16]. The M-1 phase can be produced by non-stoichiometry and pressure [23]. These facts imply the sensitivity of the non-collinear magnetic phases to the chemical composition and lattice volume. Recently, monocrystalline ANMn$_3$ films have been successfully grown on different substrates, such as SrTiO$_3$, BaTiO$_3$, and LSAT [41,42]. This opens a possibility to engineer antiperovskite compounds with nearly degenerate energies of the different magnetic phases by proper doping and suitable epitaxial strain produced by the substrate. In particular, the dynamic strain generated by a piezoelectric substrate, such as PMN-PT, can be used to realize the reversible switching between different magnetic phases. Furthermore, the energy barrier between different magnetic phases, such as $\Gamma_{5g}$ and $\Gamma_{4g}$, could be overcome by the spin transfer torque induced by a spin polarized current [26]. This would lead to a magnetic phase transition between local energy minima without lattice distortion. These possibilities make the ANMn$_3$ family of materials a promising platform for the AHE based applications of spintronic devices.

## VI. Summary

In this work, we have studied the intrinsic AHC in different non-collinear magnetic phases of GaNMn$_3$, as a representative of a broader materials family of antiperovskite compounds ANMn$_3$ (A is a main group element). Based on the symmetry analysis and first-principles DFT calculations, we showed that the nearly degenerate non-collinear AFM $\Gamma_{5g}$ and $\Gamma_{4g}$ phases of GaNMn$_3$ have zero and finite AHC, respectively. This difference was explained by the different magnetic space group symmetry of these phases. We also predicted that GaNMn$_3$, in the non-collinear ferrimagnetic M-1 phase, exhibits large AHC which is comparable to the AHC in elemental ferromagnets, such as iron.

We argued that by doping and strain it is possible to engineer the ANMn$_3$ compounds where the energy difference between these magnetic phases could be small, so that an external stimulus, such as the dynamic strain or the spin transfer torque could produce switchable magnetic phase transitions. Our work demonstrates that the antiperovskite family of non-collinear magnetic materials is a good platform to realize the multiple AHE states in a single compound, which is promising for novel spintronic applications.

**Acknowledgments**

This research was supported by the National Science Foundation (NSF) through the DMREF program (grant DMR-1629270) and the nCORE, a wholly owned subsidiary of the Semiconductor Research Corporation (SRC). Computations were performed at the University of Nebraska Holland Computing Center.

**APPENDIX**

The AHC tensor depends on geometry used in transport measurements. Table 1 above shows the AHC tensor for GaNMn$_3$ (001) growth orientation corresponding to the standard Cartesian coordinates with $x$ along [100], $y$ along [010], and $z$ along [001] directions. For GaNMn$_3$ (111) sample, the AHC can be measured for a charge current parallel to the Ga-Mn Kagome lattice. Here we show the AHC tensor for a GaNMn$_3$ (111) sample, with $x$ pointing along $[\bar{1}10]$, $y$ along $[\bar{1}\bar{1}2]$ and $z$ along [111] directions. The respective AHC tensor $\sigma_{[111]}$ can be obtained from

$$\sigma_{[111]} = R \, \sigma_{[001]} \, R^{-1} \quad (A1)$$

where $\sigma_{[001]}$ is the AHC tensor for GaNMn$_3$ (001) and R represents the respective rotation matrix. The resulting AHC tensors for $\Gamma_{4g}$ and $\Gamma_{5g}$ phases are shown in Table A1, where $\sigma'_{xy} = -68 \, \Omega^{-1} cm^{-1}$.

**Table A1:** AHC matrix tensors for $\Gamma_{5g}$ and $\Gamma_{4g}$ magnetic phases with $\hat{x}||[\bar{1}10]$, $\hat{y}||[\bar{1}\bar{1}2]$ and $\hat{z}||[111]$.

| Magnetic Phase | $\Gamma_{5g}$ | $\Gamma_{4g}$ |
|---|---|---|
| AHC tensor | $\begin{bmatrix} 0 & 0 & 0 \\ 0 & 0 & 0 \\ 0 & 0 & 0 \end{bmatrix}$ | $\begin{bmatrix} 0 & \sigma'_{xy} & 0 \\ -\sigma'_{xy} & 0 & 0 \\ 0 & 0 & 0 \end{bmatrix}$ |